\documentstyle[11pt,newpasp,twoside,epsf]{article}
\input psfig.tex

\def\edcomment#1{\iffalse\marginpar{\raggedright\sl#1\/}\else\relax\fi}
\reversemarginpar

\def\aa{{A\&A}}

\def\aj{{AJ}}
\def\al{$\alpha$}

\def\apj{{ApJ}}
\def\apjs{{ApJS}}
\def\asec{$^{\prime\prime}$}

\def\deg{$^{\circ}$}

\def\etal{{et al. }}

\def\hst{{\it HST}}

\def\lamb{$\lambda$}
\def\lax{{$\mathrel{\hbox{\rlap{\hbox{\lower4pt\hbox{$\sim$}}}\hbox{$<$}}}$}}
\def\gax{{$\mathrel{\hbox{\rlap{\hbox{\lower4pt\hbox{$\sim$}}}\hbox{$>$}}}$}}
\def\simlt{\lower.5ex\hbox{$\; \buildrel < \over \sim \;$}}
\def\simgt{\lower.5ex\hbox{$\; \buildrel > \over \sim \;$}}
\def\lum{erg s$^{-1}$}

\def\mnras{{MNRAS}}

\def\pasp{{PASP}}

\def\percm2{cm$^{-2}$}

\begin{document}
\title{The Population of AGNs in Nearby Galaxies}
 \author{Luis C. Ho}
\affil{The Observatories of the Carnegie Institution of Washington, 813 Santa 
Barbara St., Pasadena, CA 91101, U.S.A.}

\begin{abstract}
This contribution reviews the properties of nuclear activity in nearby 
galaxies, with emphasis on results obtained from current optical surveys and 
multiwavelength follow-up observations thereof.

\end{abstract}

\vspace*{-0.3cm}
\section{Introduction}

In a meeting which celebrates different techniques of surveying AGNs, it is 
appropriate to remind ourselves of the properties of the AGN population in 
nearby galaxies.  Nearby AGNs are important for at least two reasons.  First, 
they inform us of the faint end of the local ($z\approx0$) AGN luminosity
function, a fundamental constraint on a variety of statistical considerations 
of the AGN population.  Second, as the evolutionary endpoints of quasars, 
they present an opportunity to study black hole accretion in a unique regime 
of parameter space.  This paper limits itself to three topics.  Section 2 
summarizes some general statistics resulting from optical searches for 
nearby AGNs.  The interpretation of these results largely hinges on our 
understanding of the physical origin of low-ionization nuclear emission-line 
regions (LINERs; Heckman 1980), which is the subject of Section 3.  Finally, 
Section 4 draws some inferences on the demographics of massive black holes.

\vspace*{-0.3cm}
\section{Statistics of Nearby AGNs from the Palomar Survey}
Most AGN surveys rely on selection criteria that isolate some previously 
known characteristics of these objects.  Common strategies to find quasars, 
for example, employ color cuts to highlight the UV excess typically present in 
AGN spectra, or objective prism plates to identify objects with strong emission 
lines. Other wavelength-specific techniques to find candidate AGNs include 
combing areas of the sky in the radio or X-rays.  Infrared-based methods use 
temperature selection to cull sources ``warmer'' 
than might be expected for star-forming galaxies.   While all of these 
techniques have been successful, each introduces biases specific to the 
wavelength.  And all require follow-up optical spectroscopy to confirm the 
AGN identification, to classify its type, and to determine its redshift.  

A more direct, less biased approach is to spectroscopically survey 
every object within a particular region of the sky to a given optical
magnitude limit.  This, of course, is an expensive route, and in practice 
one is confined to 
%
%
%
go deep over only a small solid angle or stay relatively 
shallow over a wider area.  In addition, when covering a wide area, a 
morphological cut has to be made so that one does not waste a lot of time 
observing foreground stars.

\begin{figure}
\vbox{
\hbox{\vsize 3.0in
\hskip -0.3truein
\psfig{file=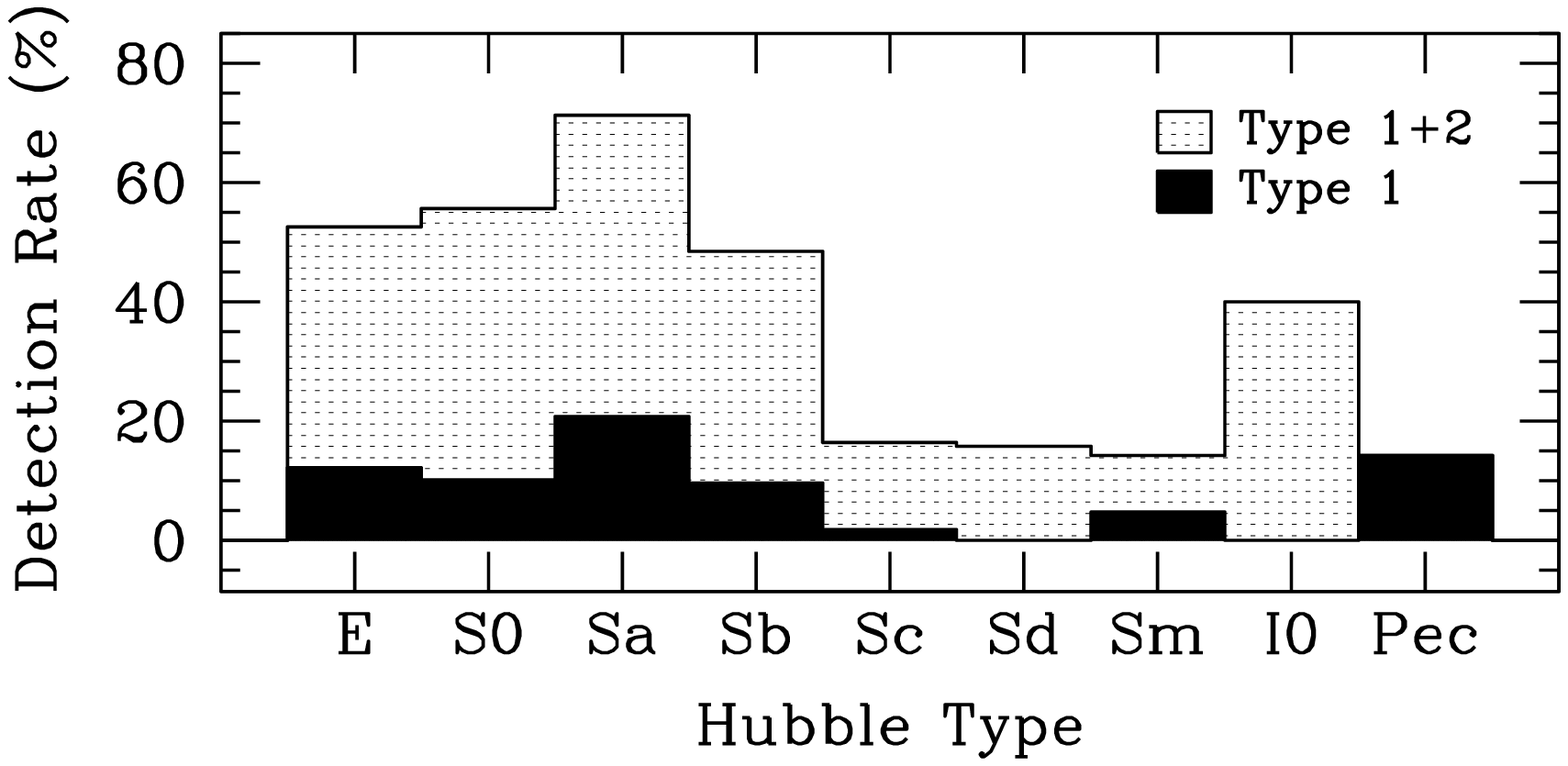,height=2.5truein,width=3.5truein,angle=0}
}
\vspace*{-1.9in}
\hskip +3.2truein
\vbox{\hsize 2.0in
\noindent{Fig. 1.} Detection rate of AGNs as a function of Hubble type in the
Palomar survey.  ``Type 1'' AGNs (those with broad H$\alpha$) are shown
separately from the total population (types 1 and 2).  Adapted from Ho et al.
(1997b).
}
}
\end{figure}

The results presented in the rest of this paper stem from the the Palomar
optical spectroscopic survey of nearby galaxies (Filippenko \& Sargent 1985; 
Ho, Filippenko \& Sargent 1995).  In brief, the Palomar 200-inch telescope was 
employed to take moderate-dispersion, high-quality spectra of 486 bright 
($B_T\,\leq$ 12.5 mag), northern ($\delta\,>$ 0\deg) galaxies,
with the primary aim of conducting an accurate census of the AGN population in 
the nearby ($z\,\approx 0$) universe.  The Palomar survey, the most sensitive
of its kind (see Ho 1996 for a comparison with previous optical studies),
produced a comprehensive, homogeneous catalog of spectral classifications of
nearby galaxies (Ho et al. 1997a, 1997b, 1997c).  The
main results of the survey are the following.
(1) AGNs are very common in nearby galaxies (Fig. 1).  At least 40\% of all
galaxies brighter than $B_T$ = 12.5 mag emit AGN-like spectra.  The
emission-line nuclei are classified as Seyferts, LINERs, or transition objects 
(LINER/H~II), and most have very low luminosities compared to traditionally
studied AGNs.  The luminosities of the H$\alpha$ emission line range from
10$^{37}$ to 10$^{41}$ erg s$^{-1}$, with a median value of $\sim$10$^{39}$
erg s$^{-1}$.
(2) The detectability of AGNs depends strongly on the morphological type of
the galaxy, being most common in early-type systems (E--Sbc).
The detection rate of AGNs reaches 50\%--75\% in ellipticals, lenticulars,
and bulge-dominated spirals but drops to \lax 20\% in galaxies classified as
Sc or later.
(3) LINERs make up the bulk (2/3) of the AGN population and a sizable
fraction (1/3) of all galaxies.
(4) A significant number of objects show a faint, broad
(FWHM $\approx$ 1000--4000 km s$^{-1}$) H$\alpha$ emission line that
 qualitatively resembles emission arising from the conventional broad-line
region of ``classical'' Seyfert 1 nuclei and quasars.

\vspace*{-0.3cm}
\section{The Nature of LINERs\footnote{Note that this 
paper is concerned only with compact, {\it nuclear} LINERs ($r$ \lax\ 200 pc), 
which are most relevant to the AGN issue.  LINER-like spectra are 
often also observed in extended nebulae such as those associated with 
cooling flows, nuclear outflows, and circumnuclear disks.}}

If LINERs are powered by a nonstellar central source, then they clearly would 
be the most abundant type of AGNs in nearby galaxies.  However, ever since 
their discovery, the physical origin of LINERs has been controversial.
The AGN 

\begin{figure}
\vbox{
\hbox{\vsize 3.0in
\hskip -0.1truein
\psfig{file=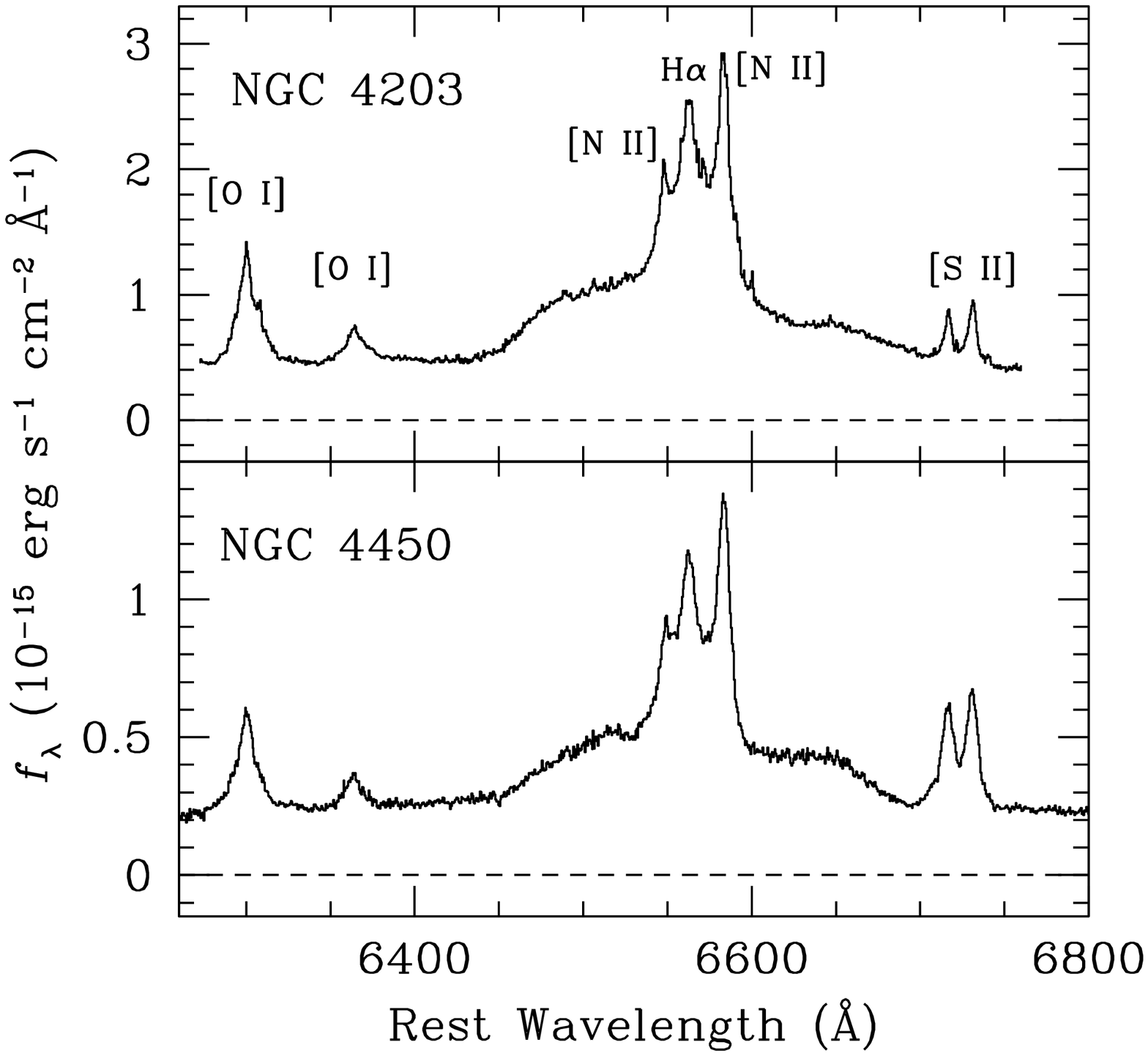,height=3.5truein,angle=0}
}
\vspace*{-1.2in}
\hskip +3.7truein
\vbox{\hsize 1.5in
\noindent{Fig. 2.} Double-peaked broad H\al\ lines detected in STIS spectra
of NGC 4203 (Shields et al. 2000) and NGC 4450 (Ho et al. 2000).
}
}
\end{figure}

\noindent
interpretation for LINERs is only one of several possible explanations, 
as recently reviewed by Barth (2001).  The most uncertain 
subset are the type~2 LINERs and transition objects, whose energetics may 
arise, at least in part, from stellar processes.  

Ho (1999a) used several lines of evidence to argue that a significant fraction 
of LINERs are genuine AGNs.  Some of these are repeated below, along with 
updated information when available.

\vspace*{0.1cm}
\noindent
{\it (1) The host galaxies of LINERs are similar to those of Seyfert nuclei.}\ 
Both classes reside in bulge-dominated hosts (E--Sbc), quite distinctive from 
star-forming H~II nuclei, which are found mainly in Hubble types Sc and 
later.  To the extent that bulges invariably contain massive black holes 
(e.g., Magorrian et al. 1998; Kormendy \& Gebhardt 2001; R. Green, these 
proceedings), this is consistent with the idea that LINERs are associated with 
accretion processes.

\vspace*{0.1cm}
\noindent
{\it (2) Some of the best candidates for massive black holes are in LINER 
galaxies.}\  Well-known examples include M81, M84, M87, and the Sombrero 
galaxy.  The rapidly growing list of objects with kinematically determined 
black hole masses (see, e.g., Kormendy \& Gebhardt 2001; Ho 2001b) continues 
to support this.

\vspace*{0.1cm}
\noindent
{\it (3) LINERs contain broad-line regions.}\  The Palomar survey discovered 
that 15\%--25\% of the LINER population are ``type~1'' LINERs --- LINERs with 
a directly visible broad component of H$\alpha$ emission (Ho et al. 1997c).  
The broad-line component, however, is generally rather weak; the measurement
is extremely challenging, requiring careful subtraction of the underlying 
starlight and deblending of complex line profiles.  The robustness of the 
broad H\al\ detections, therefore, remained to be evaluated by independent 
observations.  Rix et al. (2002) recently used STIS on \hst\ to obtain nuclear 
spectra of a statistically complete subsample of emission-line nuclei selected 
from the original Palomar survey.  The small aperture (0\farcs2) used in these 
observations excludes 

\begin{figure}
\vbox{
\hbox{
\hskip 0truein
\psfig{file=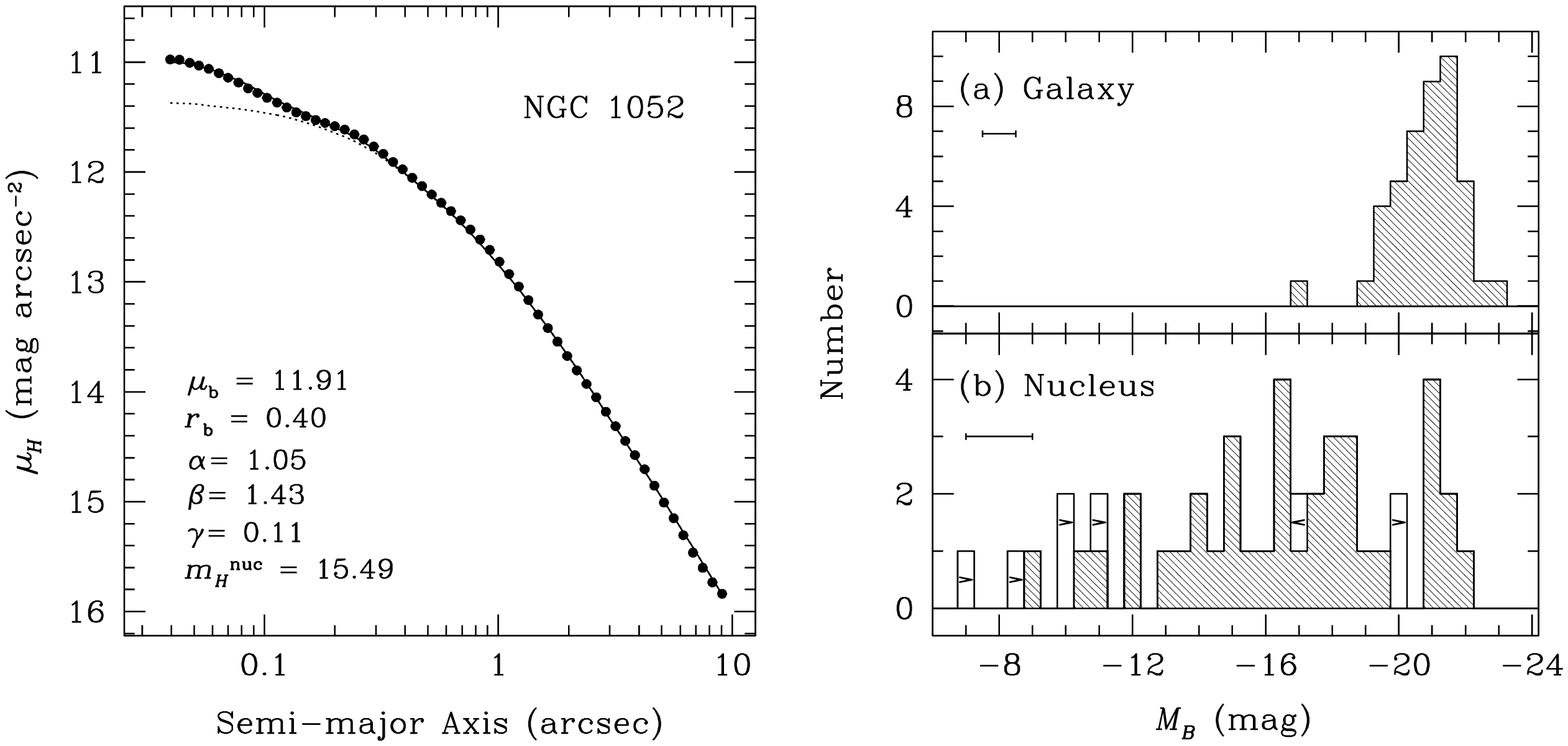,height=2.5truein,angle=270}
}
}
\vskip +1.1cm
Fig. 3. ({\it Left}) NICMOS/{\it HST}\ $H$-band central surface brightness
profile of NGC 1052.  The central point source is extracted after modeling
the underlying bulge light.  Adapted from Ravindranath et al. (2001).
({\it Right}) Optical ($B$) absolute magnitudes for Seyfert~1 galaxies. The
top panel plots the integrated emission from the galaxy, and the bottom panel
shows the light from the nucleus alone, as determined from {\it HST}\
photometry.  Adapted from Ho \& Peng (2001).
\end{figure}

\vspace*{-0.8cm}
\noindent
most of the contaminating bulge light and 
enables a much more straightforward assessment of the presence of broad lines. 
These data, and other related observations using \hst, have shown 
that the statistics of broad H\al\ emission derived from the Palomar survey 
are essentially robust (Rix et al. 2002).  

However, the \hst\ spectra revealed that in some objects (Fig.~2) the broad 
line does show a qualitative difference in the form of a double-peaked, 
asymmetric profile, highly reminiscent of similar structures found in some 
broad-line radio galaxies that are commonly interpreted as signatures of a 
relativistic accretion disk (e.g., Chen, Halpern, \& Filippenko 1989).  
Double-peaked broad lines are supposed to be quite rare; according to 
Eracleous \& Halpern (1994), they are seen in only 10\% of radio-loud AGNs 
($\sim$1\% of all bright AGNs).  Given the relatively high detection frequency 
of these objects in serendipitous \hst\ observations (M81: Bower et al. 1996; 
NGC 4203: Shields et al. 2000; NGC 4450: Ho et al. 2000; NGC 4579: Barth 
et al. 2001), we suspect that this phenomenon is much more common than 
previously thought.  Indeed, Ho et al. (2000) discuss how double-peaked broad 
lines might arise naturally in nearby nuclei with low-accreting black holes.   

\vspace*{0.1cm}
\noindent
{\it (4) Compact nuclei are common in LINERs.}\   But they are not easy to 
measure.  Emission from the host galaxy invariably swamps the nuclear 
component.  The contrast problem is not as severe at some wavelengths compared 
to others (e.g., UV), but generally it never goes away.  In order to reliably 
quantify the nuclear fluxes from these sources, one needs observations not 
only with good sensitivity but also high angular resolution.  The second point 
is crucial: generally angular resolutions of \lax 1\asec\ are necessary to 
unambiguously disentable a tiny central core embedded in a bright bulge.  
The requisite data, although in principle not that challenging to obtain with
modern facilities, have been subject to systematic analysis only quite
recently.  

Figure~3 illustrates the practical challenge.  The left panel shows the 
central surface brightness profile of the well-studied LINER NGC 1052 measured 
in the $H$ band with NICMOS on \hst\ (Ravindranath et al. 2001).  The central 
point source is faint, but quite well defined at $r$ \lax\ 0\farcs1.  It almost 
certainly would have escaped notice in typical ground-based images. Even with 
\hst\ resolution, however, note that it is not trivial to extract the 
point-source magnitude unambiguously in the presence of a cuspy bulge profile.  
The right panel in the figure generalizes this problem to a well-defined sample 
of nearby Seyfert 1 galaxies recently studied by Ho \& Peng (2001).  It shows 
that although the contrast problem is most acute in LINERs, it is nonetheless 
quite important as well in Seyfert galaxies.  The top panel, which plots the 
integrated optical ($B$) absolute magnitudes of the entire galaxy, gives the 
familiar result that Seyfert galaxies typically have $M_B\,\approx\,-21$ mag, 
roughly corresponding to $L^*$ (e.g., Ho et al. 1997b).  On the other hand, 
the true magnitude of the Seyfert {\it nucleus}\ ---  the quantity most 
pertinent to the AGN phenomenon and most analogous to observations of quasars 
--- ranges from $M_B\,\approx\,-21$ to $-9$, a factor up to $10^5$ fainter than 
the integrated luminosity.  Clearly, any quantitative discussion of the 
physical properties of low-luminosity AGNs, be they LINERs or Seyferts, would 
be meaningless if this were not taken into account.

The incidence of compact nuclei in LINER galaxies depends on wavelength, 
but in general the detection rate is quite high, on the order of 50\% or 
greater. The most robust statistics come from radio and X-ray observations, 
since these bands are least affected by dust obscuration.  

Several recent studies have exploited the high angular resolution of the 
VLA to search for compact radio cores in the Palomar galaxies (Van Dyk \& Ho 
1998; Nagar et al. 2000; Filho, Barthel, \& Ho 2000; Ho \& Ulvestad 2001). The 
VLA can deliver fairly sensitive (rms $\sim$ 50--100 $\mu$Jy) radio continuum 
images efficiently in ``snapshot'' mode.  In general the observations have 
been done at 6, 3.6, or 2~cm, with resolutions of $\sim$1\asec.  The main 
results are: (a) compact cores are  quite common, being present in 
$\sim$50\%--80\% of the objects; (b) they are weak, typically 
$P_{\rm 6cm}\,\approx\,10^{18}-10^{21}$ W Hz$^{-1}$; (c) a sizable fraction 
of them have flat or even inverted spectra; and (d) they have relatively simple 
structures, usually well described by a single unresolved core. Many sources 
remain unresolved when examined at milli-arcsecond resolution using VLBI 
techniques (e.g., Wrobel, Fassnacht, \& Ho 2001; Filho, Barthel, \& Ho 2001; 
Ulvestad \& Ho 2002, in preparation), although some show jetlike linear extensions 
(Falcke et al. 2000).

The Palomar galaxies have also been the subject of intense scrutiny in the 
X-rays, both in the soft-energy band using {\it ROSAT}\ (Komossa, B\"ohringer, 
\& Huchra 1999; Roberts \& Warwick 2000; Halderson et al. 2001) and in the 
hard-energy band using {\it ASCA}\ (Terashima, Ho, \& Ptak 2000; Terashima et 
al. 2001).  The coarse angular resolution of these satellites, however, 
severely limits one's ability to reliably measure the weak signal from the 
nucleus.  This is the domain of {\it Chandra}. A large, well-defined subsample
of the Palomar galaxies is currently being imaged with ACIS.  The preliminary
findings, reported by Ho et al. (2001), suggest that compact X-ray sources 
astrometrically coincident with the radio

\begin{figure}
\vbox{
\hbox{
\hskip 0truein
\psfig{file=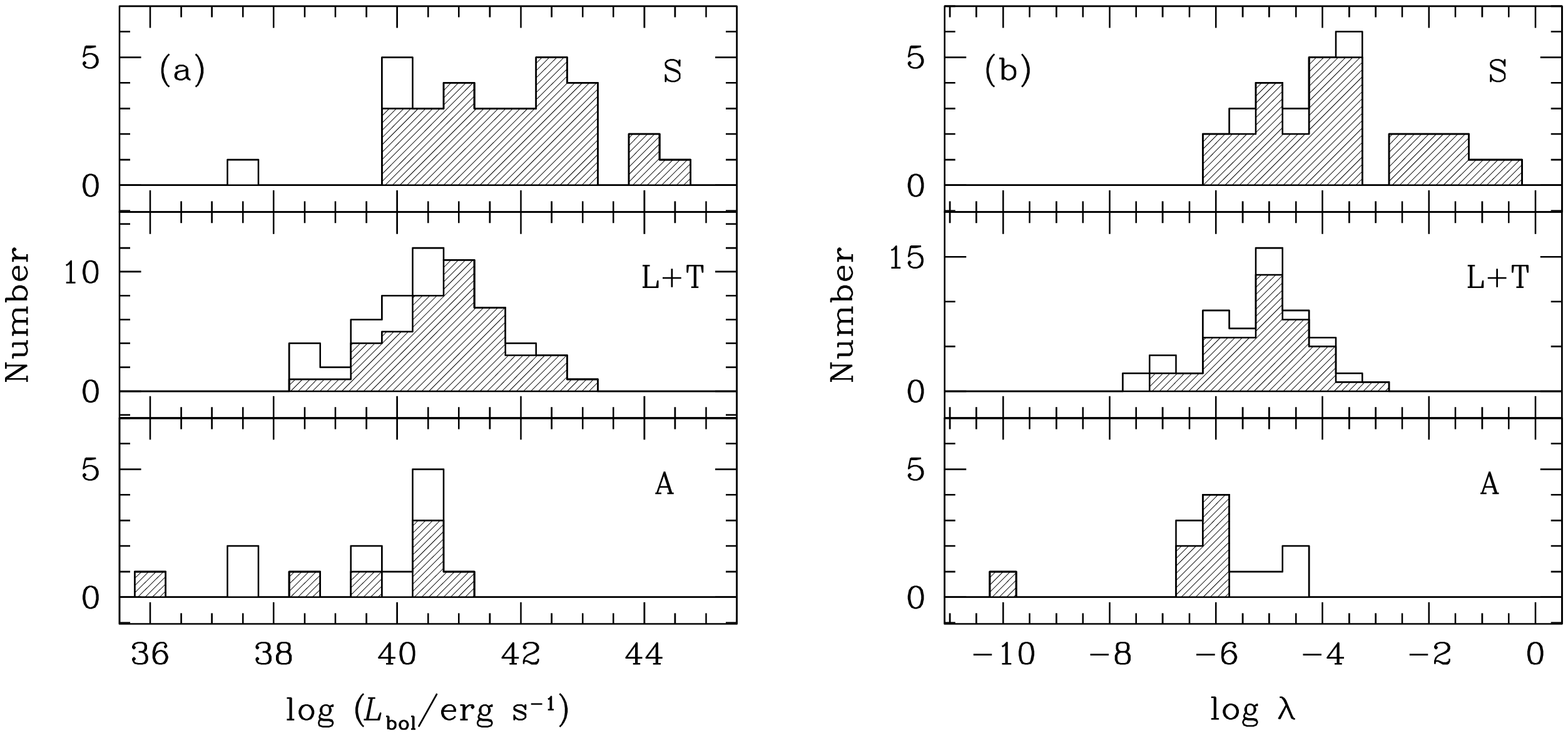,height=2.5truein,angle=270}
}
}
\vskip +1.0cm
Fig. 4.  Distribution of ({\it a}) nuclear bolometric luminosities and
({\it b}) Eddington ratios $\lambda\,\equiv\,L_{\rm bol}/L_{\rm Edd}$ for
Seyferts (S), LINERs and transition objects (L+T), and absorption-line nuclei
(A).  Open histograms denote upper limits.  Adapted from Ho (2002, 
in preparation).
\end{figure}

\vspace*{-0.4cm}
\noindent
cores are again commonplace.  They are detected universally in 
type~1 LINERs and Seyferts, relatively often in type~2 LINERs and Seyferts, 
but very rarely in transition nuclei.  These statistics need to be verified 
with the full sample, which is forthcoming (Ho et al. 2002, in preparation).

\vspace*{0.1cm}
\noindent
{\it (5) LINERs are highly sub-Eddington systems.}\  Recent advances in 
black hole mass determinations in nearby galaxies gives us an opportunity to 
compare the nuclear luminosities described above with the corresponding 
Eddington luminosities.  Ho (2002, in preparation) estimated nuclear 
bolometric luminosities 
for $\sim$100 Palomar galaxies based on X-ray measurements.  As shown in 
Figure~4{\it a}, the majority of the objects have $L_{\rm bol}$ \lax\ 
$10^{42}$ \lum, with the interesting trend that Seyferts are systematically 
more luminous than LINERs (including transition objects), which in turn are 
brighter than absorption-line nuclei (objects with no nuclear optical emission 
lines).  More revealing still is the trend with 
$\lambda\,\equiv\,L_{\rm bol}/L_{\rm Edd}$ (Fig.~4{\it b}).  The distributions 
of \lamb\ systematically shift to lower values following the sequence 
S $\rightarrow$ L+T $\rightarrow$ A.  Although LINERs and Seyferts broadly 
overlap, note that {\it all}\ LINERs are characterized by 
\lamb\ \lax\ $10^{-3}$.  

\vspace*{0.1cm}
\noindent
{\it (6) The spectral energy distributions (SEDs) are peculiar.}\  
Specifically, the SEDs generically lack the optical-UV ``big blue bump'' 
(Ho 1999b; Ho et al. 2000), a near-universal feature of unobscured 
high-luminosity AGNs usually attributed to thermal emission from an optically 
thick, geometrically thin accretion disk. Another attribute of the SEDs of 
low-luminosity AGNs, especially of LINERs, is that they are typically ``radio 
loud,'' defined here by the convention that the radio-to-optical luminosity 
ratio exceed some fiducial value, say $R > 10$, as normally adopted in quasar 
studies.  In fact, Ho (2001b) finds that among the $\sim$40 nearby galaxies 
with kinematically determined black hole masses, essentially 

\begin{figure}
\vbox{
\hbox{\vsize 2.7in
\hskip -0.1truein
\psfig{file=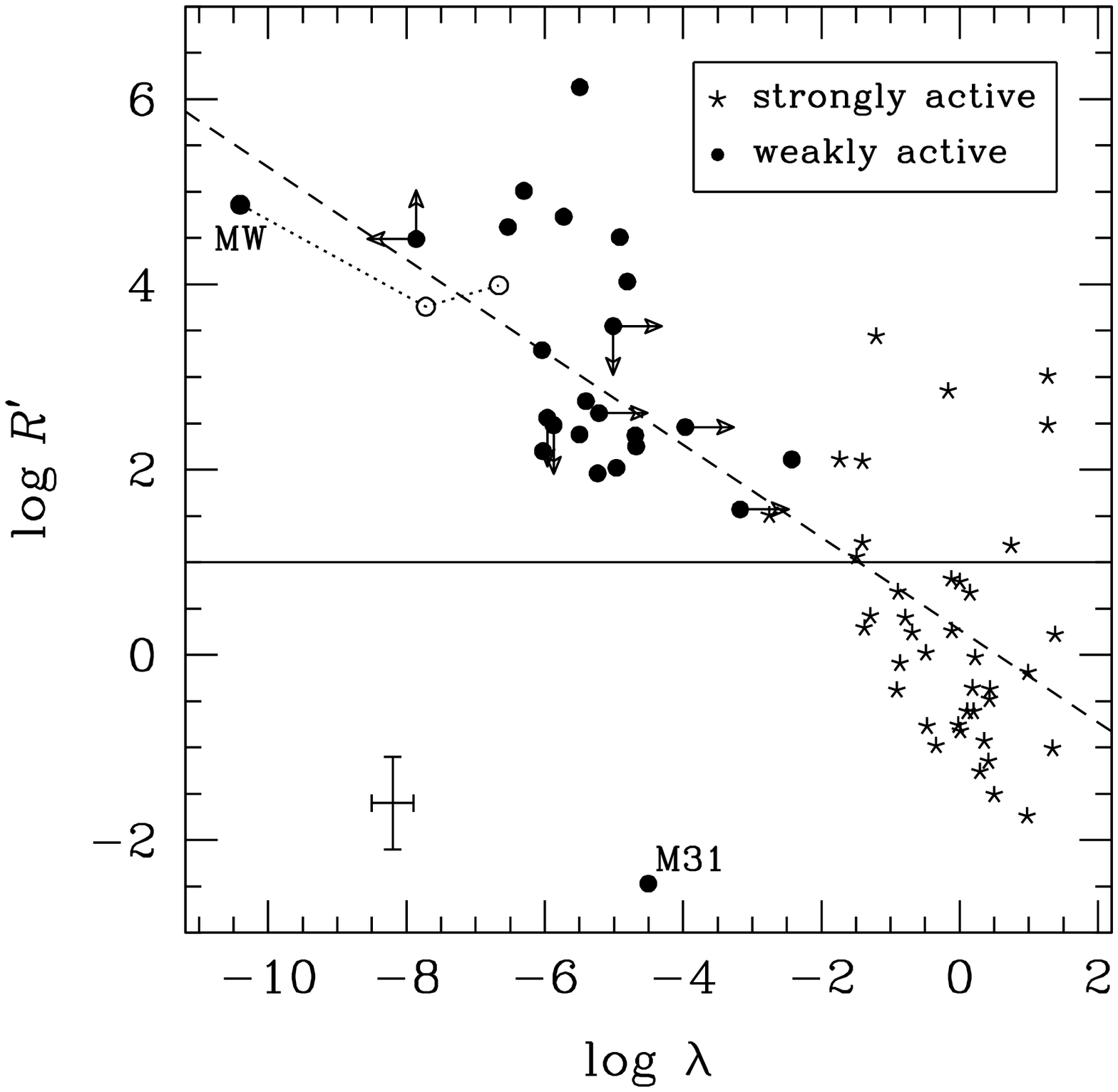,height=3.2truein,angle=0}
}
\vspace*{-2.4in}
\hskip +3.4truein
\vbox{\hsize 1.8in
\noindent{Fig. 5.} Distribution of the nuclear radio-to-optical luminosity
ratio $R^{\prime}$ vs.  $\lambda\,\equiv\,L_{\rm bol}/L_{\rm Edd}$. The
{\it solid line}\ marks the formal division between radio-loud and radio-quiet
objects, $R^{\prime}$ = 10.  The {\it dashed line}\ is the best-fitting linear
regression line.  Adapted from Ho (2001b).
}
}
\end{figure}

\vspace*{0.0cm}

\noindent
{\it all}\ are 
radio loud (Fig.~5).  Since many of these objects are spiral galaxies, this 
result challenges the conventional wisdom that radio loudness is an attribute 
unique to giant ellipticals (see Ho \& Peng 2001 for a related discussion 
in the context of Seyfert galaxies).  Moreover, Figure~5 shows that the 
degree of radio loudness evidently increases systematically with decreasing 
accretion rate, here parameterized by \lamb.  

\vspace*{0.1cm}
\noindent
{\it (7) LINERs have nonstandard central engines.}\  As further elaborated in 
Ho et al. (2000) and Ho (2001a), LINERs share several characteristics expected 
of massive black holes fed by an advection-dominated accretion flow (see reviews 
by Narayan, Mahadevan, \& Quataert 1998; Quataert 2001) instead of a canonical 
thin accretion disk.  The most pertinent of these are (a) the low Eddington 
ratios and the inferred low mass accretion rates, (b) the absence of the big 
blue bump in the SEDs, (c) the characteristic radio loudness and its dependence 
on $L_{\rm bol}/L_{\rm Edd}$, (d) the prevalance of double-peaked broad 
emission lines and the disk structure they imply, and (e) the low-ionization 
state of the line-emitting gas, a possible consequence of the particular form 
of the SEDs.

\vspace*{0.1cm}
\noindent
{\it (8) The nature of type~2 LINERs and transition objects is uncertain.}\
An important unsolved problem is what fraction of the narrow-lined sources
should be considered genuine AGNs.  To be sure, AGN-like LINER~2s do exist.
Barth, Filippenko, \& Moran (1999a) discovered that the LINER~2 nucleus of 
NGC 1052 shows prominent broad-line emission when viewed in scattered light. 
The unification scheme, popular for Seyfert galaxies, evidently applies to at 
least some LINERs.  If one assumes that all LINERs strictly adhere to the 
simplest form of the unified model (that all type~2 sources are 
simply obscured type~1 sources), and that the ratio of LINER~2s to LINER~1s 
is the same as that of Seyfert~2s to Seyfert~1s in the Palomar survey, then 
the true AGN fraction among all LINERs is estimated to be between 45\% and
65\%, depending on whether transitions objects are excluded or included,
respectively (Ho 1996, 1999a).  This estimate may be overly optimistic for
at least two reasons.  First, the spectropolarimetric observations of these
very faint sources are demanding and still quite limited 

\begin{figure}
\vbox{
\hbox{\vsize 2.7in
\hskip -0.1truein
\psfig{file=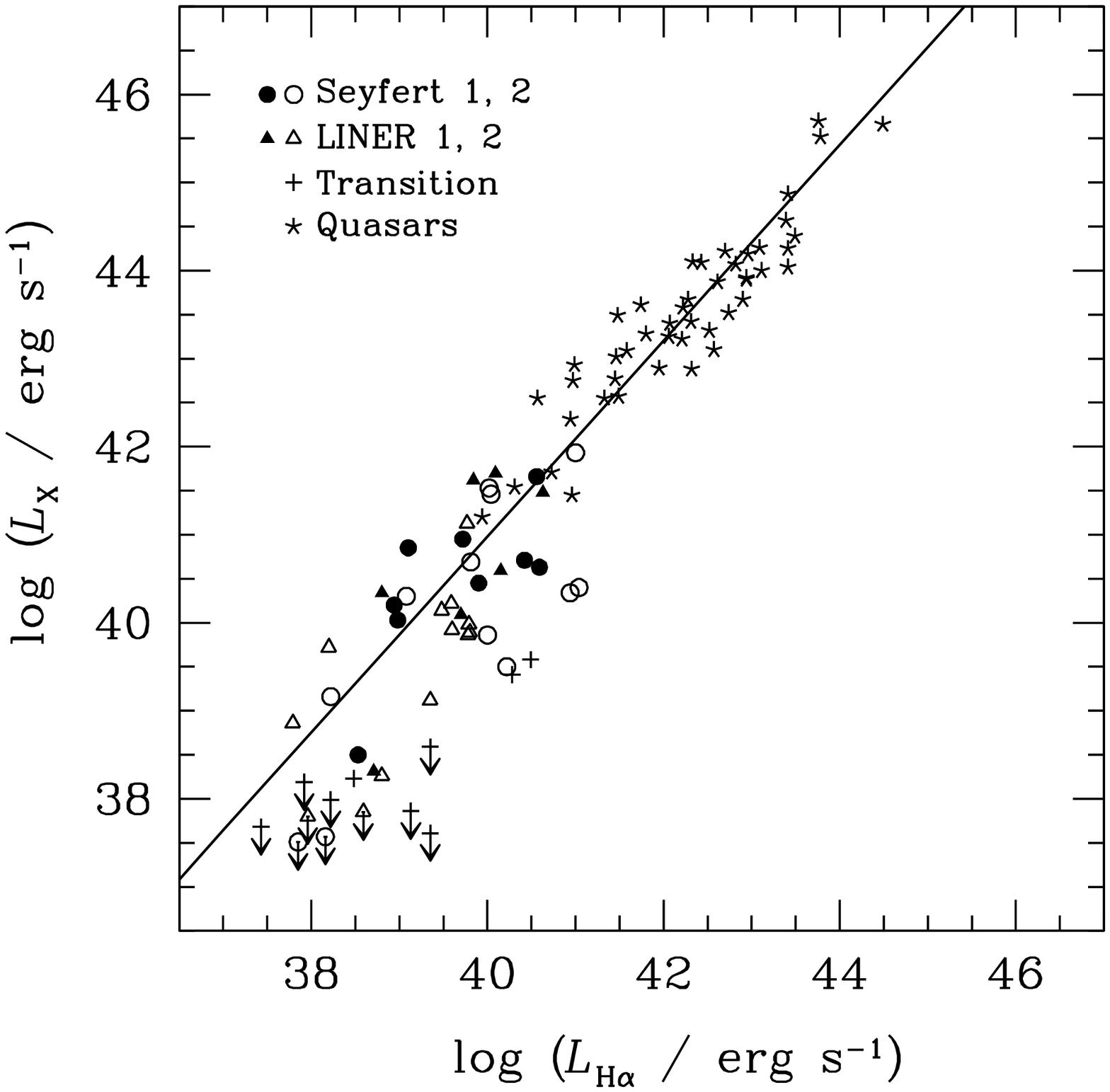,height=3.2truein,angle=0}
}
\vspace*{-1.4in}
\hskip +3.4truein
\vbox{\hsize 1.8in
\noindent{Fig. 6.}
Correlation between 2--10 keV X-ray luminosity and H\al\ luminosity for AGNs
of various types.  Adapted from Ho et al. (2001).
}
}
\end{figure}

\vspace*{+0.0cm}
\noindent
(Barth et al. 1999a, 
1999b), and so it may be unwise to generalize these results prematurely.  
And second, young, massive stars have been found to contribute significantly 
to the energy budget of some narrow-lined low-ionization objects (e.g., Maoz 
et al. 1998; Barth \& Shields 2000).  The first caveat may be partly offset by 
the fact that some genuine low-luminosity AGNs (e.g., M84, M87, Sombrero) may 
{\it intrinsically}\ lack a broad-line region (Barth 2001).  

The resolution of this problem requires access to data which are sensitive 
to a spectral region that is universally present in AGNs and is relatively 
unaffected by dust obscuration or modest photoelectric absorption.  To be 
effective, the data must have high angular resolution; to survey large 
samples, necessary for statistical confidence, the data need to be gathered 
in an efficient mode.  As described above (item 4), the current generation 
of radio and hard X-ray data both meet these requirements; we give greater 
weight to the latter because it carries a larger fraction of the AGN bolometric 
luminosity.  Figure~6, taken from the preliminary results of the 
{\it Chandra}/ACIS survey of Ho et al. (2001), illustrates the potential 
of the X-ray data.  Type~1 AGNs (LINERs, low-luminosity Seyferts, classical 
Seyferts, and quasars) obey a well-defined, linear correlation between hard 
X-ray ($2-10$ keV) and H\al\ luminosity, down to $L_{\rm X}$(2--10~keV) 
$\approx$ $3 \times 10^{38}$ \lum. The $L_{\rm X}-L_{{\rm H}\alpha}$ 
correlation strongly supports the hypothesis that all of these sources, which 
span $\sim$8 orders of magnitude in luminosity, share the same physical origin. 

Note that the majority of the type~2 sources (LINERs and Seyferts) {\it do}\
have compact X-ray cores.  They follow the type~1 objects in Figure~6, albeit 
with somewhat greater scatter and offset toward lower 
$L_{\rm X}/L_{{\rm H}\alpha}$ by about a factor of 10.  Interestingly, the 
X-ray hardness ratios of these sources suggest that absorption is not the 
culprit for their X-ray weakness.  The transition objects, on the other hand, 
appear distinctively different as a class: most are undetected in X-rays, with 
upper limits as low as $L_{\rm X}$(2--10~keV) $\approx$ $3 \times 10^{37}$ 
\lum.  The incidence of compact radio cores in these objects (Filho et al. 
2000) also appears to be lower than in LINERs or Seyferts.

\vspace*{-0.2cm}
\section{Implications for Black Hole Demography}

To the extent that an AGN signature signifies accretion onto a massive black 
hole, a local AGN census gives us a lower limit on the fraction of nearby 
galaxies hosting massive black holes.  If we accept that all LINERs, transition 
objects, and Seyferts are genuine AGNs, then the AGN statistics from the 
Palomar survey (Fig.~1) imply that black holes exist in $>$40\% of all galaxies 
with $B_T\,\leq$ 12.5 mag.  For bulge-dominated systems (E--Sbc), this fraction 
climbs to $>$50\%--75\% --- not inconsistent with the 100\% claimed by direct 
kinematic studies.  But how confident are we about the AGN-ness of nearby 
emission-line nuclei?  The status of objects classified as Seyferts does not
seem to be in dispute (for reasons that reflect historical bias than 
concrete evidence).  In any case, LINERs make up the majority of the 
population, and so deservingly receive closer scrutiny.  The evidence 
presented in \S~3 shows that type~1 LINERs almost surely must be AGNs.  
The verdict for type~2 LINERs is much less clear-cut at the moment.  While 
some sources have been shown to be powered by hot stars, others clearly 
have all the hallmark attributes of bona fide AGNs.  Perhaps the most telling 
statistic is the incidence of compact radio and hard X-ray cores; although 
not yet well quantified, the preliminary results indicate that the detection 
rate is quite high in both bands (\gax 50\%).  On the other hand, the 
``transition objects,'' postulated to be composite LINER/H~II systems, 
may very well turn out to be unrelated to AGNs.  Or if present, the AGN 
component must be significantly weaker than in ``regular'' LINERs.  
If we conservatively exclude all transition objects from the AGN pool, 
then the overall AGN fraction should be reduced by 13\% (Ho et al. 1997b).


\acknowledgments{
L.C.H. acknowledges financial support through NASA grants from the Space 
Telescope Science Institute (operated by AURA, Inc., under NASA contract 
NAS5-26555).}

\end{document}